\documentclass[prl,superscriptaddress,twocolumn]{revtex4}

\usepackage{ulem}
\usepackage[dvips]{graphicx}%
\usepackage{bm,color} 
\usepackage{amsmath,amssymb}
\newcommand{\braket}[2]{\langle #1 | #2 \rangle}
\newcommand{\ket}[1]{\left |  #1 \right \rangle}
\newcommand{\bra}[1]{ \left \langle #1  \right |}
\newcommand{\ave}[1]{ \langle #1   \rangle}

\newcommand{\ketbra}[2]{\ket{#1} \bra{#2}}

\begin{document}
\title{Discrete Fourier-based Correlations for Entanglement Detection}

\author{Ryo Namiki}\affiliation{Department of Physics, Graduate School of Science, Kyoto University, Kyoto 606-8502, Japan}
\author{Yuuki Tokunaga}\affiliation{NTT Secure Platform Laboratories, NTT Corporation, 3-9-11 Midori-cho, Musashino, Tokyo, 180-8585, Japan}\affiliation{Japan Science and Technology Agency, CREST, 5 Sanban-cho, Chiyoda-ku, Tokyo, 102-0075, Japan}

\begin{abstract} 
We introduce two forms of correlations  on two $d$-level (qudit) systems for entanglement detection. The correlations can be measured via experimentally tractable two local measurement settings and their separable bounds are determined by discrete Fourier-based uncertainty relations. They are useful to estimate lower bounds of the Schmidt number in order to clarify generation of a genuine qudit entanglement.  We also present inseparable conditions for multi-qudit systems associated with the qudit stabilizer formalism as another role of the correlations on the inseparability problem.   
\end{abstract}

\maketitle

%

 Entanglement plays a central role in the development of quantum physics and quantum information science. In order to understand the entanglement phenomena, it is important to ask what is the simplest test for entanglement. While there are several approaches on this question \cite{Au06,Guh02,Toth05L}, an experimentally plausible simple test would be formulated from the correlations of the measurement outcomes of two local non-commutable observables. A reasonable choice of the pair observables is complementary ones whose eigenstates form two conjugate bases 
   related by a Fourier transform with each other.
 The correlation of the canonical position and momentum operators is employed for the detection of the continuous-variable entanglement \cite{Duan-Simon,Reid89}. 
Similar formalism for discrete systems is described in terms of two Pauli operators.

For the two $2$-level (\textit{qubit}) systems denoted by $A$ and $B$, one of the simplest conditions for the existence of entanglement might be 
\cite{Toth05,Au06,Guh07,Toth05L}
\begin{eqnarray}
\langle \hat Z_A\otimes  \hat Z_B + \hat X_A\otimes \hat X_B \rangle > 1 \label{XZ},
\end{eqnarray} where $\hat Z := \ket{0}\bra{0}-\ket{1} \bra{1}$ and $\hat X := \ket{0}\bra{1} +\ket{1}\bra{0}$ are  the Pauli $Z$ and $X$ operators with the fixed $Z$-basis $\{|0\rangle ,\ket{1}\}$. Associated with the standard quantum key distribution protocol, Eq. (\ref{XZ}) corresponds to the condition that the average of the bit error rate and phase error rate is lower than 25\% \cite{Curt04}, namely, it is sufficient to clarify the inseparability if the average correlation of the $Z$-basis bits and $X$-basis bits exceeds 75\%: 
\begin{eqnarray}
\frac{1}{2}\sum_{j=0,1} \langle  \ket{j}\bra{j}\otimes \ket{j}\bra{j}  + \ket{\bar j}\bra{\bar j}\otimes \ket{\bar j}\bra{\bar j}  \rangle > \frac{3}{4}, \label{biterror}
\end{eqnarray}
where the $X$-basis is given by the Fourier transform of the $Z$-basis as $\ket{\bar 0} = (\ket{0}+\ket{1})/\sqrt 2$ and $\ket{\bar 1} = (\ket{0}-\ket{1})/\sqrt 2 $.  In order to test this condition, two local measurement settings of the measurement in the $Z_A Z_B$-basis and the measurement in the $X_A X_B$-basis are sufficient. Inequalities (\ref{XZ}) and (\ref{biterror}) suggest that the existence of strong correlations both in the complementary aspects can be a signature of entanglement. Such types of coexisting correlations are also significant  for quantification of entanglement in discrete-variable systems \cite{Koashi07} as well as in continuous-variable systems  \cite{Gie03}.

  There has been growing interest in generating entanglement beyond on the qubit pairs and in verifying that the generated states possess genuinely multi-dimensional entanglement   {\cite{Mai01,Tokunaga08, Inoue09,Li11}}. 
 To eliminate the possibility of lower dimensional entanglement  the Schmidt number  \cite{Ter20} has been utilized in the experiments \cite{Tokunaga08,Inoue09}. The   \textit{Schmidt number} is defined by the minimum of the maximum Schmidt-rank of the pure states to construct the density operator. It indicates the number of degrees of freedom contributing to the entanglement and is an entanglement monotone to quantify the entanglement \cite{Sper11}. 
An ultimate goal of the experiments is to generate the maximally entangled state (MES). The fidelity to the MES is called the \textit{MES fraction}. It represents the closeness of the generated entangled state to  the MES.
 Hence, a major theoretical step is to provide accessible tools for estimating the fraction of MES and for evaluating the entanglement without tomographic effort {\cite{Li11}}. %

Another interesting topic is detection of entanglement for multipartite systems where Eq. (\ref{XZ}) has  already constituted as an entanglement test in the stabilizer formalism \cite{Toth05,Toth05L,Guh07}. However, the entanglement tests associated with the multi-dimension stabilizer formalism \cite{Zhou03,Gott} have not been established yet.

In this Letter we generalize the conditions of Eqs. (\ref{XZ}) and (\ref{biterror}) for two $d$-level ($qudit$) systems by using  
 two different forms of the discrete Fourier-based uncertainty relations \cite{Mass08, Lars90}. We also show that the coexisting correlations provide lower bounds of the  MES fraction and Schmidt number for detection of  a genuine multi-dimensional entanglement. We further find two other inseparable conditions on multipartite systems with the stabilizer operators for Greenberger-Horne-Zeilinger (GHZ) states and for cluster states on the multi-$d$-level systems.  The correlations can be measured by two local measurement settings and are thought to be useful in   the experiments.

Let us define the generalized Pauli $Z$ and $X$ operators as
$\hat Z: = \sum_{j= 0}^{d-1} e^{ i\omega j} |j  \rangle\langle j|$ and 
$ \hat X: = \sum_{j=0}^{d-1}|j+1 \rangle\langle j|$,   
 assuming a fixed $Z$-basis $\{ \ket{0},\ket{1}, \cdots, \ket{d-1}\}$ with modulo-$d$ conditions $|j+ d\rangle = |j\rangle $ and $\omega  := 2 \pi /d $.
They satisfy 
$\hat X^d =\hat Z^d  =  \openone$ and $\hat Z^m \hat X^l  = e^{ i\omega  l m } \hat X^l \hat  Z^m$. 
For the qubit system ($d =2$), the operators $\hat Z= \hat Z^\dagger$ and $\hat X= \hat X^\dagger$ correspond to the ordinary Pauli operators. %
For the cases of $d>2$, $\hat Z$ and $\hat X$ are not self-adjoint but still unitary operators. 
We define the $X$-basis as a Fourier transform of the $Z$-basis by 
$| \overline {k} \rangle :=  \hat Z ^ k  \left(
\frac{1}{\sqrt d} \sum_{j=0}^{d-1 }\ket{j } \right) =\hat Z ^ k|\overline {0} \rangle$. 
 The $X$-basis gives the diagonal expression of 
 $\hat X = \sum_{j= 0}^{d-1} e^{ - i\omega j} |\bar j  \rangle\langle\bar j|$.
We can verify $ | \langle j | \overline {k} \rangle| = 1/\sqrt d $ for any $j$ and $k$. The bases which satisfy this condition are called \textit{mutually unbiased}.  

\textit{Discrete Fourier-based uncertainty relations.---}
 In order to recall two uncertainty relations \cite{Mass08, Lars90} associated with the Fourier-based relations, let us write the probability distributions for $Z$-basis $P(j) :=\langle j| \rho |j  \rangle$ and for $X$-basis $ \bar { P} (j) :=\langle \overline{j}| \rho  |\overline{j}  \rangle $. For any two probability distributions we have the relation $\sum_{j} ( P^2(j) +\bar P^2 (j)) \le 2$. For the distributions of two mutually unbiased bases, a strict condition holds: 
  \cite{Lars90,Luis07}:
\begin{eqnarray}
\sum_{j=0}^{d-1} ( P ^2(j) +  \bar P ^2 (j) )  \le 1 +\frac{1}{d}. \label{URH2}
\end{eqnarray}

There is another uncertainty relation concerning the complex-valued expectation value of the unitary operators \cite{Op95,Mass08}.
The action of the unitary operator $\hat Z$ ($\hat X ^\dagger$) applies the phase factor $e^{i\omega j }$ onto the $j$-th eigenstate $\ket{j}$ ($\ket{\bar j}$), and its expectation value can be seen as the complex amplitude.  
 The relation  $\langle j |  \hat Z \rho | j \rangle =P(j) e^{i \omega  j } $ implies that $\langle \hat Z \rangle $ maps the probability distribution of the $ Z$-basis, $\{  P(j) \} $,  to a complex number inside the regular $d$-gon inscribed in the unit circle on the complex plane. Similarly, the relation $\langle \hat X \rangle =\sum_j \bar P (j) e^{-i \omega  j } $ maps the $X$-basis distribution $\{  \bar P (j) \} $ to a complex number inside the regular $d$-gon inscribed in the unit circle. {While the unitary condition of $\hat Z$ and $\hat X$ implies $|\langle \hat Z \rangle|\le 1 $, $|\langle \hat X \rangle|\le 1 $ and $|\langle \hat Z \rangle |+ |\langle \hat X \rangle  |\le 2 $, the Fourier-based relation  on $\{  P(j) \} $ and $\{ \bar  P(j) \} $ suggests definite trade-off between  the complex amplitudes $\langle \hat Z \rangle $ and $\langle \hat X \rangle $. 
  Intuitively, the amplitudes  $\langle \hat Z \rangle $ and $\langle \hat X \rangle $ cannot be found on the unit circle, simultaneously.}  A tight condition due to the incompatibility is given by 
 Theorem 2 of Ref. \cite{Mass08}  
\begin{eqnarray}
 & &     | \langle Z \rangle  | \cos \theta    + | \langle  X  \rangle  | \sin \theta      \nonumber \\ 
&\le& \frac{1}{2}\|( Z+ Z^\dagger) \cos \theta  +( X+X^\dagger ) \sin \theta \|  =:  \|\hat \chi_\theta   \|  , \label{UR08}
\end{eqnarray}
where $\theta \in [0, \pi /2 ]$ and $\|\hat O  \| := \max_{\braket{u}{u}=1}  |\bra{u} \hat O \ket{u}|$.

\textit{Inseparable conditions from the uncertainty relations.---} 
We define the MES on a two-qudit system as 
\begin{eqnarray}
|\Phi_{0,0} \rangle :&=& \frac{1}{\sqrt d }\sum_{j=0}^{d-1}    |j  \rangle_A | j  \rangle _B  = 
 \frac{1}{\sqrt d }\sum_{j=0}^{d-1}    |\overline {  j}  \rangle_A |\overline{ -j}  \rangle _B,   \label{def-SMES}
\end{eqnarray}
and  define the total correlation on the Fourier bases as a generalization of the operator in Eq. (\ref{biterror}), 
\begin{eqnarray}
 \hat C_d &:= & \sum_{j=0}^{d-1}\left(\ket{j}\bra{j} \otimes\ket{j}\bra{j}  +\ket{\overline j}\bra{\overline j} \otimes\ket{\overline {-j}}\bra{\overline {- j}} \right) . \label{deftotal}\end{eqnarray}
From Eq. (\ref{def-SMES}) it is direct to see that the MES $|\Phi_{0,0} \rangle$ is a simultaneous eigenstate of the projective operators: 
\begin{eqnarray}
\sum_{j=0}^{d-1}(\ket{j}\bra{j} \otimes\ket{j}\bra{j} ) |\Phi_{0,0}\rangle &=& |\Phi_{0,0} \rangle , \nonumber \\  
\sum_{j=0}^{d-1}(\ket{\overline j}\bra{\overline j} \otimes\ket{\overline {-j}}\bra{\overline {- j}} )  |\Phi_{0,0}\rangle &=&|\Phi_{0,0}\rangle  . 
\end{eqnarray} 
This implies that the MES provides perfectly correlated $Z$-basis outcomes and perfectly anti-correlated $X$-basis outcomes, simultaneously.  
  Hence, the central question is how high one can enlarge the total correlation without entanglement.

 For the product state $\ket{\phi}\ket{\varphi}$, we can estimate
$\ave{\hat C_d}_{\phi \otimes \varphi }  = \sum_{j=0}^{d-1} (  P_{\phi}(j)  P_{\varphi}(j) + \overline P_{\phi}(j)  \overline P_{\varphi}(- j) ) 
  \le \sqrt{\sum_{j=0}^{d-1}  ( P_{\phi}^2(j) +{\overline P_{\phi}}^2(j)  )}   \sqrt{\sum_{j=0}^{d-1} (  P_{\varphi}^2(j)  +  {\overline P_{\varphi}}^2(- j) )}  \le   1 + {1}/{d}$ 
where the first inequality comes from Schwarz inequality $|\Vec a| |\Vec b|  \ge |\Vec a\cdot \Vec  b| $.  
Then, the trivial relation $\sum_{j=0}^{d-1}  {\overline P_{\varphi}}^2(- j) = \sum_{j=0}^{d-1}  {\overline P_{\varphi}}^2(j)$ and Eq. (\ref{URH2}) lead to the final expression. 
Consequently, for the separable state $\rho_s = \sum_{i} p_i \ket{\phi_i}\bra{\phi_i } \otimes \ket{ \varphi_i }\bra{ \varphi_i  }$, we have 
$\ave{\hat C_d}_{\rho_s}= \sum_i p_i \ave{\hat C_d}_{\phi_i \otimes \varphi_i }  \le  \sum_i p_i \left ( 1 +\frac{1}{d}\right )=   1 +\frac{1}{d}$. 
 The equality is achieved by, e.g., the product state $\ket{\phi}\ket{\varphi}=\ket{0}\ket{0}$. 
We thus obtain the inseparable condition for the total correlation $\hat C_d$ 
 as $d$-dimensional counterpart of Eq. (\ref{biterror}): 
\begin{eqnarray}
  \sum_{j=0}^{d-1} \ave{ \ket{j}\bra{j} \otimes\ket{j}\bra{j}  +\ket{\overline j}\bra{\overline j} \otimes\ket{\overline {-j}}\bra{\overline {- j}} }  
>    1 +\frac{1}{d}. \label{ex1} \end{eqnarray} 
{This condition is equivalent to the condition in Ref. \cite{Li11} for two-qudit case with $l=2$
and can also be derived from the method of Ref. \cite{Hof03} by considering the sum of the variances $\sum_j [\delta^2 ( \ket{j}\bra{j}_A -  \ket{j}\bra{j}_B ) + \delta^2 (\ket{\overline { j}}\bra{\overline { j}}_A  - \ket{\overline {-j}}\bra{\overline {- j}}_B ) ]$  
 with  Eq. (\ref{URH2}) and binding the non-linear terms $\sum_j [(P_A (j) - P_B (j ))^2+ (\overline{P}_A (j) -  \overline{P}_B ( - j ))^2 ]\ge 0 $. }

\begin{figure}[htbp]
\includegraphics[width= \linewidth]{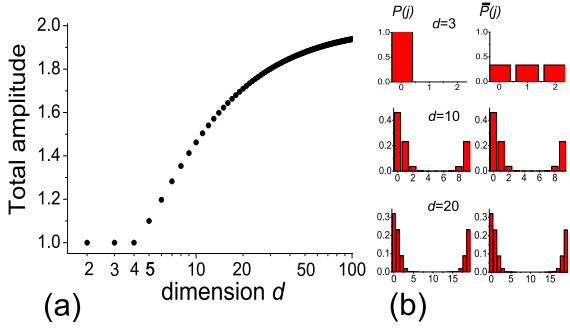}
\caption{(Color online). (a) The separable boundary of the total amplitude $M_d =\max_{ \phi  } \left[   | \langle\hat  Z \rangle_\phi   |^2   + | \langle  \hat X  \rangle_\phi    | ^2  \right] $ as a function of the local state-space  dimension $d$. (b) The set of optimal distributions that attain the maximum for $d= 3$, $10$, and $20$. \label{fig3}}
\end{figure}
Similarly we can find another inseparable condition 
 in terms of the generalized Pauli operators as follows.
From Eq. (\ref{def-SMES}) it  is also direct to see that the MES is a simultaneous eigenstate of the relative amplitudes of the generalized Pauli $ Z$s and $X$s as  $(\hat Z_A - \hat Z_B)|\Phi_{0,0}\rangle  =  0 $ and 
$(\hat X_A - \hat X_B^\dagger )|\Phi_{0,0}\rangle  = 0$. 
By multiplying $(\hat Z_A -\hat Z_B)^\dagger $ and $(\hat X_A -\hat X_B^\dagger )^\dagger $ from left side respectively 
 we have the relations with the self-adjoint operators,
\begin{eqnarray}
 {(\hat Z_A \hat Z_B^\dagger + \hat Z_A^\dagger \hat Z_B})  |\Phi_{0,0}\rangle &=& 2 |\Phi_{0,0}\rangle ,  \nonumber \\  
( \hat X_A \hat X_B +\hat X_A ^\dagger \hat X_B ^\dagger  )|\Phi_{0,0}\rangle &=&2|\Phi_{0,0}\rangle . \label{zzxx}
\end{eqnarray} 
Let us define the total correlation of the amplitudes in correspondence with Eq. (\ref{XZ})
as \begin{eqnarray}
  \hat R_{d}  := \frac{1}{2} \left( \hat Z_A\hat Z_B^\dagger + \hat Z_A^\dagger \hat  Z_B \right)+ \frac{1}{2} \left(  \hat  X_A\hat  X_B +\hat X_A ^\dagger \hat   X_B ^\dagger \right ) . \label{AmpCorr} \end{eqnarray} 
Then, Eq. (\ref{zzxx}) implies  
$\ave{ \hat R_{d} }_{\Phi_{0,0}}=2$. 
This is different from the situation that $ | \langle \hat  X \rangle |= 1$ and $| \langle  \hat  Z\rangle |=1$ are simultaneously satisfied. Actually, single expectation values vanish, $\langle \hat  X \rangle = \langle  \hat  Z\rangle=0$. 
It seems impossible to attain $\langle \hat R_{d}\rangle =2$ by the separable states  
although either $\langle  \hat Z_A\hat Z_B^\dagger + \hat Z_A^\dagger \hat Z_B \rangle =2 $ or $\langle \hat X_A \hat X_B +\hat X_A ^\dagger \hat  X_B ^\dagger  \rangle  =2 $ can be attained by the product state, $|0\rangle |0 \rangle$ or $|\bar 0\rangle | \bar 0 \rangle$,  respectively. %
Hence, the question is how large 
 the expectation value of the total amplitude $\langle \hat R_{d}\rangle $ can be without entanglement.  

For the product state, we can find a form of the upper bound on the total amplitude with Schwarz inequality as
\begin{eqnarray}
2\ave{\hat R_{d} }_{\phi \otimes \varphi} &= & \ave{\hat Z_A}_\phi\ave{\hat Z_B^\dagger}_\varphi  + \ave{\hat Z_A^\dagger}_\phi  \ave{\hat Z_B}_\varphi \nonumber \\ 
& & +     \ave{ \hat  X_A}_\phi \ave{\hat X_B }_\varphi +  \ave{\hat X_A ^\dagger}_\phi \ave{\hat X_B ^\dagger }_\varphi  \nonumber \\ 
&\le  &2 \left(   | \langle \hat Z_A   \rangle_\phi | | \langle \hat Z_B^\dagger \rangle_\varphi |   + | \langle\hat   X_A \rangle_\phi | | \langle\hat  X_B \rangle_ \varphi |   \right)  \nonumber \\ 
&\le & 2 \sqrt{ | \langle \hat  X \rangle_\phi |^2 + | \langle \hat Z \rangle_\phi |^2   } \sqrt{   | \langle \hat  X  \rangle_ \varphi |^2 +| \langle \hat Z ^\dagger \rangle_\varphi |^2 }   \nonumber \\ 
&\le    & 2 \max_{ \phi  } \left[   | \langle\hat  Z \rangle_\phi   |^2   + | \langle  \hat X  \rangle_\phi    | ^2    \right]  =:2 M_d.  \label{ineq2}
\end{eqnarray}
For the separable state, this inequality implies 
$\ave{\hat R_{d} }_{\rho_s}  =   \sum_i p_i \ave{\hat R_{d} }_{\phi_i \otimes \varphi_i} \le  \sum_i p_i  M_d =M_d$. 
Therefore, we obtain another inseparable condition as $d$-dimensional  counterpart of Eq.  (\ref{XZ}): 
\begin{eqnarray}
\frac{1}{2}\ave{\hat Z_A\hat Z_B^\dagger + \hat Z_A^\dagger \hat  Z_B +\hat  X_A\hat  X_B +\hat X_A ^\dagger \hat   X_B ^\dagger}>M_d. \label{ex2}
\end{eqnarray}
We can write 
$M_d =   \max_{ \phi  } \left[   | \langle\hat  Z \rangle_\phi   |^2   + | \langle  \hat X  \rangle_\phi    | ^2  \right]  = \left(\max _{0\le \theta \le \frac{\pi}{2} } \|\hat \chi_ \theta  \|  \right)^2$    
by using the uncertainty relation of Eq. (\ref{UR08}) and the relation $\sqrt{   |  \alpha |^2   + | \beta  | ^2 } = \max_{\theta } \left( |\alpha  | \cos \theta    + | \beta | \sin \theta   \right)$  with   $\alpha = \langle \hat Z\rangle$ and $\beta = \langle \hat X\rangle$. Inequality (\ref{ineq2}) is saturated by the product of the local state that picks up the maximal eigenvalue of the operator $\hat \chi_\theta $ at the optimal value of $\theta$.  
Figure \ref{fig3} shows $M_d $ from $d=2$ to $d=100$ and three sets of the two Fourier distributions relevant for the optimal local state. The optimal value of $\theta $ is  $\theta = \pi/4$ except for $d = 3$ where $M_d=1$ is achieved at $\theta = 0$, or $\pi/2$.
{Note that  we can also obtain  inseparable conditions with the convex sum of the operators $p(\hat Z_A \hat Z_B^\dagger+\hat Z_A^\dagger \hat Z_B) +(1-p)( \hat X_A \hat X_B+\hat X_A^\dagger  \hat X_B^\dagger)$ in that case we set $\alpha  =\sqrt p \langle \hat Z\rangle$ and $ \beta =\sqrt{1- p} \langle \hat X\rangle$. {It might be topical that the two inseparable conditions of Eqs. (\ref{ex1}) and (\ref{ex2}) coincide  for $d=2$ and $d=3$ whereas  
 the sets of the detectable entangled states include mutually exclusive subsets  {
  for $d \ge 4 $ as in FIG.  \ref{supfig3}. An element of the subsets can be found by considering the noisy MESs $ \hat \psi (p) = p \hat \Phi_{0,0}+(1-p) \hat \Phi_{\lfloor \frac{d}{2} \rfloor, \lfloor \frac{d}{2} \rfloor}$ and $ \hat \phi (p) = p \hat \Phi_{0,0}+(1-p) \hat \Phi_{1,0}$, 
  where we write  $\hat \Phi_{l,m} =\ket{ \Phi_{l,m}}\bra{ \Phi_{l,m}}$ with  the  Bell states $|\Phi_{l,m} \rangle_{AB} := \hat X_A^l \hat Z_B^m  |\Phi_{0,0} \rangle_{AB} $ and the floor function $\lfloor \cdot \rfloor$. 
  }

\begin{figure}[htbp]\includegraphics[width=0.9\linewidth]{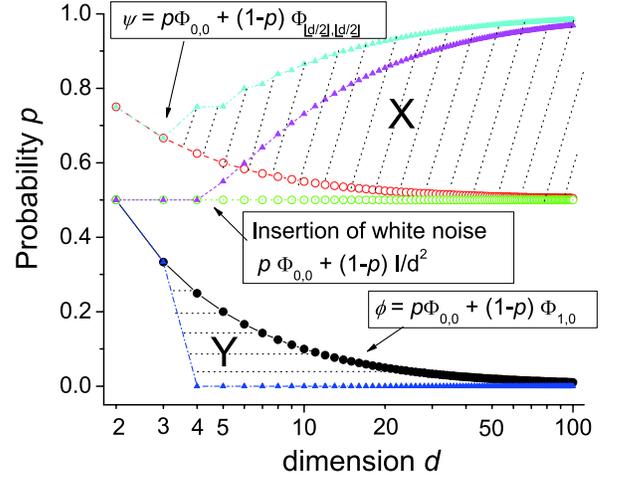}\caption{(Color online). Noise tolerance threshold $p$ for the states $ \hat \psi (p)$, $ \hat \phi (p)$, and $p  \hat \Phi_{0,0} +(1-p) \openone /{d^2}$. Circles are for the qudit correlations of Eq. (\ref{ex1}) and triangles are for the amplitude correlations of Eq. (\ref{ex2}). 
 For $d \ge 4$, there exist parameter regime of $p$ denoted by the symbol X (Y) where $ \hat \psi (p)$ [$ \hat \phi (p)$] is shown to be entangled due to Eq. (\ref{ex1}) [Eq. (\ref{ex2})] but its inseparability cannot be verified by Eq. (\ref{ex2}) [Eq. (\ref{ex1})].    \label{supfig3} }\end{figure}
\textit{MES fraction and Schmidt number.---}
We proceed to the stronger statement for genuine  qudit entanglement detection.  
Interestingly, the two operators take the Bell-diagonal forms:
$\hat C_d = \sum_{l=0}^{d-1}  \hat \Phi_{l,0}+\sum_{m=0}^{d-1}  \hat \Phi_{0,m}$ and $\hat R_d =  \sum_{l,m=0}^{d-1}\left\{(\cos l\omega +\cos m\omega ) \hat \Phi_{l,m} \right\}$.  
Using the completeness relation $\sum_{l,m=0}^{d-1}  \hat \Phi_{l,m}= \openone$, we can confirm 
$\hat C_d   =  \openone + \hat \Phi_{0 ,0} -\sum_{l,m=1}^{d-1}  \hat \Phi_{l,m} \le  \openone + \hat \Phi_{0 ,0}$ and  
$ \hat R_d  \le    (1- \cos \omega ) \hat \Phi_{0,0} +  (1+ \cos \omega ) \sum_{l,m=0}^{d-1} \hat \Phi_{l,m}  =   (1- \cos \omega ) \hat \Phi_{0,0} + (1+ \cos \omega )  \openone$.  
We thus obtain two lower bounds of the MES fraction as
\begin{eqnarray}\ave{\hat \Phi_{0 ,0} }&\ge&  \ave{\hat C_d} -1, \nonumber \\
\ave{\hat \Phi_{0 ,0} }& \ge & \frac{1}{1- \cos \omega }\left(\ave{\hat R_d}-(1+ \cos \omega ) \right) \nonumber . \end{eqnarray} 
Note that the state is uniquely determined to be $\ket{\Phi_{0 ,0}}$ in the case of  $\langle \hat C_d \rangle =2 $ or in the case of $\langle \hat R_d \rangle =2 $. {From Lemma 1 of \cite{Ter20},} if the fraction is $\langle \hat \Phi_{0 ,0} \rangle > (k-1)/d$, the Schmidt number is at least $k$ {(See also, Eq. (6) of \cite{San01})}. Therefore, the Schmidt number is at least $k$ if either $\langle \hat C_d \rangle > 1+(k-1)/d $ or $\langle \hat R_d \rangle > [(d-k+1)\cos \omega +(d+k-1)] /d $ is satisfied.
Hence, by  setting $k=d$, we obtain the criterion for generation of the genuine qudit entanglement with the Schmidt number of $d$. One can also calculate a lower bound of Entanglement of Formation from the lower bounds of the  MES fraction \cite{Ter00a}.  Therefore, the coexisting complementary correlations have been linked to both the detection and quantification of entanglement.

For the experimental tests, the qudit correlation $\ave{\hat C_d}$ of Eq.  (\ref{deftotal}) requires the total number of $2 d $ expectation values for the rank-one projections $\{\ket{j}\bra{j} \otimes \ket{j}\bra{j} \} $ and $\{\ket{\overline j}\bra{\overline j} \otimes \ket{\overline{ -j}}\bra{\overline{ -j} } \} $. On the other hand, the amplitude correlation $\ave{\hat  R_d}$ of Eq. (\ref{AmpCorr}) requires the total number of $2d^2$ expectation values for the rank-one projections $\{\ket{j}\bra{j} \otimes \ket{k}\bra{k} \} $ and $\{\ket{\overline j}\bra{\overline j} \otimes \ket{\overline{ k}}\bra{\overline{k} } \} $. All elements can be obtained by the measurements of the joint $Z$-basis, $\{\ket{j}\ket{k}\} $, and the joint  $X$-basis, $\{ \ket{\bar j}\ket{ \bar k}\} $. A standard implementation could be the combination of the fixed basis measurement and the local unitary transformation that induces the Fourier transform. It might be instructive to emphasize that $\ave{\hat R_d}$ is real-valued by showing the decomposition with the projections: $ \hat R_d = \sum_{j,k}  \{ \cos [\omega (j-k)] \ketbra{j}{j}  \otimes \ketbra{k}{k}+ \cos [\omega (j+k)] \ketbra{\bar j}{\bar j}  \otimes \ketbra{\bar k}{\bar k}   \}$.

\textit{Entanglement test with stabilizer operators.---} 
Finally, we show the conditions for the entanglement detection associated with the stabilizer formalism with qudits \cite{Gott,Zhou03}.
 Let us define the stabilizer operators for $N$-qudit GHZ state $\ket{G}=\frac{1}{\sqrt{d}} \sum_{k=0}^{d-1}\ket{k}^{\otimes N}$  as $S_1:=\Pi_{k=1}^{N} \hat X_k $ and 
$S_m:=\hat Z_{m-1} \hat Z_{m}^\dagger$
for $ m=2,3,\cdots, N$. 
The stabilizer conditions  $S_k \ket{ G} = \ket{G}$ and  $S_k^\dagger \ket{ G} = \ket{G}$ hold.
 For the product states, similar to Eq. (\ref{ineq2}), 
 we have
$\langle  S_1 + S_1^\dagger \rangle  + \langle  S_m + S_m ^\dagger \rangle =\langle \hat X_1 \rangle \langle \hat X_2 \rangle \cdots \langle \hat X_N \rangle + \langle \hat X_1^\dagger  \rangle \langle \hat X_2^\dagger \rangle \cdots \langle \hat X_N^\dagger \rangle + \langle \hat Z_{m-1} \rangle\langle \hat Z_m^\dagger  \rangle + \langle \hat Z_{m-1}^\dagger \rangle\langle \hat Z_m  \rangle \le 2 (|\langle \hat X_{m-1} \rangle| |\langle \hat X_m \rangle| +|\langle \hat Z_{m-1} \rangle | |\langle \hat Z_m \rangle | ) \le 2M_d $. Then the same bound holds for the full separable states, and we have the condition for entanglement 
$\frac{1}{2}|\langle S_1 + S_1^\dagger \rangle  + \langle  S_m + S_m ^\dagger \rangle | > M_d$.
 This is the multipartite counterpart of Eq. (\ref{ex2}) and corresponds to a $d$-dimensional extension of Theorem 1 in Ref. \cite{Toth05} (also Eq. (11) of Ref. \cite{Toth05L}). The two measurement settings of the $N$-joint $Z$-basis and $N$-joint $X$-basis \cite{Toth05L} are sufficient to demonstrate the experimental test. 
Similarly, defining the stabilizer operators for $N$-qudit cluster state as  
 $T _1:= \hat X_1^\dagger  \hat Z_2$,
$T_m:=\hat Z_{m-1}\hat X_m^\dagger  \hat Z_{m+1}$
for $ m=2,3,\cdots, N-1 $, and $ T_{N}:=\hat X_{N-1}^\dagger  \hat Z_{N}$  \cite{Zhou03},
 we have
$\langle T_{m-1} +T_{m-1}^\dagger \rangle  + \langle T_{m } +T_{m } ^\dagger \rangle  \le 2 (|\langle \hat X_{m-1}^\dagger \rangle| |\langle \hat Z_{m } \rangle| +|\langle \hat Z_{m-1}  \rangle | |\langle \hat X_{m }^\dagger \rangle | ) \le 2M_d $ for the product states as before. This implies another inseparable condition:
$\frac{1}{2}|\langle T_{m-1} +T_{m-1}^\dagger \rangle  + \langle T_{m } +T_{m } ^\dagger \rangle | > M_d$. This is a qudit extension of Eq. (12) in Ref. \cite{Toth05L}. The two measurement settings of the $N$-joint alternate-$Z$-and-$X$ bases are sufficient for the experimental test.
{
 Because of the link between conditions of Eq. (\ref{ex1}) and Ref. \cite{Li11}, our stabilizer method has an advantage to detect the states close to $\hat \phi (p)$ over the method of Ref.  \cite{Li11}.} 

In conclusion, we have established a general relation between the discrete Fourier-based correlations and entanglement on two-qudit system. It links two local measurement settings with the basic concepts on the study of entanglement, and the findings could be useful standard tools to analyze multi-qudit entanglement.

{
 } 
\end{document}